\documentclass[12pt]{iopart}
\usepackage{graphicx}

\newcommand{\ket}[1]{\ensuremath{\left|#1\right\rangle}}
\input{Qcircuit}
\begin{document}
\title[Proposed optical realization of a two photon, four-qubit entangled $\chi$ state]{Proposed optical realization of a two photon, four-qubit entangled $\chi$ state}
\author{Atirach Ritboon, Sarah Croke and Stephen M Barnett}
\address{School of Physics and Astronomy, University of Glasgow, Glasgow G12 8QQ, UK}
\ead{a.ritboon.1@research.gla.ac.uk}
\begin{abstract}
The four-qubit states $\ket{\chi^{ij}}$, exhibiting genuinely multi-partite entanglement have been shown to have many interesting properties and
have been suggested for novel applications in quantum information processing. In this work we propose a simple quantum circuit and its corresponding optical embodiment with which to prepare photon pairs in the $\ket{\chi^{ij}}$ states.
Our approach uses hyper-entangled photon pairs, produced by the type-I spontaneous parametric down-conversion (SPDC) process in two contiguous nonlinear crystals, together with a set of simple linear-optical transformations. Our photon pairs are maximally hyper-entangled in both their polarisation and orbital angular momentum (OAM). After one of these daughter photons passes through our optical setup, we obtain photon pairs in the hyper-entangled state $\ket{\chi^{00}}$, and the $\ket{\chi^{ij}}$ states can be achieved by further simple transformations.
\end{abstract}
\submitto{\JOPT}
\maketitle
\section{Introduction}\label{sec_1}
Quantum entanglement is known to be at the heart of quantum computation and quantum information, in which it is the fundamental resource of many information processing tasks including quantum dense coding, quantum cryptography and quantum teleportation~\cite{Nielsen,Stig,Vedral,Kaye,QIbook,Bennett_1992, Ekert_1991, Bennett_1993}. It is also required for
violation of the famous Bell inequality \cite{Bellbook} and other demonstrations of quantum nonlocality \cite{Redhead}.   While bipartite entanglement is rather well understood~\cite{Peres_1996, Hill_1997, Wootters_1998}, the properties and characteristics of the various types of multipartite entanglement remain a topic of active research~\cite{Verstraete_2002}. 
In particular, there is much research to show that multipartite entanglement is helpfully employed in several quantum communication protocols, including universal error correction~\cite{Shor_1995}, quantum secret sharing~\cite{Hillery_1999}, telecloning~\cite{Murao_1999}, and deterministic secure quantum communication~\cite{Xiu_2009}. 
As one might expect, increasing the number of entangled particles leads to stronger and more dramatic demonstrations of nonlocality, or we could say that entangling greater numbers of particles leads to a wider range of nonclassical effects
that can be observed~\cite{Ardehali_1992, Lu_2014}. There are two well-known classes of genuinely tripartite entangled states: the Greenberger-Horne-Zeilinger (GHZ) state
\begin{equation}\label{eq_1}
\ket{\rm{GHZ}}=\frac{1}{\sqrt{2}}(\ket{000}+\ket{111})_{abc},
\end{equation}
and the Werner (W) state
\begin{equation}\label{eq_2}
\ket{\rm{W}}=\frac{1}{\sqrt{3}}(\ket{001}+\ket{010}+\ket{100})_{abc} .
\end{equation}
These are inequivalent under stochastic local operations and classical communication (SLOCC), which means they cannot be converted to each other under SLOCC operations~\cite{Dur_2000}. For this reason, each of these has distinct
entanglement properties. With the GHZ state, the Bell-type inequality is maximally violated~\cite{Mermin_1990, Batle_2016}, and it can be employed for open-destination teleportation by using the protocol of A. Karlsson and M. Bourennane~\cite{Karlsson_1998}. On the other hand, and in contrast to the GHZ state, losing one of particles in the W state does not make its reduced state separable~\cite{Dur_2001}. Each of these may be extended to more than three particles in a natural way:
\numparts
\begin{eqnarray}
\ket{\rm GHZ}_N &=& \frac{1}{\sqrt2} \left( \ket{0}^{\otimes N} + \ket{1}^{\otimes N} \right), \\
\ket{W}_N &=& \frac{1}{\sqrt{N}} \sum_{i=1}^N \ket{0}^{\otimes (i-1)} \ket{1} \ket{0}^{\otimes (N-i)}.
\end{eqnarray}
\endnumparts

In 2006 Yeo and Chua proposed a new type of four-qubit entangled state~\cite{Yeo_2006}
\numparts\begin{eqnarray}\label{eq_3}
\ket{\chi^{00}}=\frac{1}{\sqrt{2}}(\ket{\zeta^{0}}+\ket{\zeta^{1}})_{abcd},\label{eq_3a}\\
\ket{\chi^{ij}}=\sigma^{i}\otimes\sigma^{j}\otimes I \otimes I\ket{\chi^{00}},\label{eq_3b}
\end{eqnarray}
with
\begin{eqnarray}
\ket{\zeta^{0}}\equiv\frac{1}{2}(\ket{0000}-\ket{0011}-\ket{0101}+\ket{0110}),\label{eq_3c}\\
\ket{\zeta^{1}}\equiv\frac{1}{2}(\ket{1001}+\ket{1010}+\ket{1100}+\ket{1111}),\label{eq_3d}
\end{eqnarray}
\endnumparts
where $\sigma^{j}$ are the Pauli matrices and $I = \sigma^0$ is the identity operator.  These states cannot be transformed into the four-qubit forms of the GHZ and W states by SLOCC and so form a distinct class of entangled states. In fact the state $\ket{\chi^{00}}$ appeared also in an earlier study by Lee {\it et al} of entanglement teleportation
\cite{Lee_2002}, which showed that this state can be produced by a nonlocal transformation of the tensor product of EPR states~: $\ket{\Phi^+}_{a^\prime b^\prime}\otimes\ket{\Phi^+}_{c^\prime d^\prime}$, where $\ket{\Phi^+}=(\ket{00}+\ket{11})/\sqrt{2}$~\cite{QIbook}. Unlike other classes of multipartite entangled states, the entanglement of these states does not originate from the entanglement between any particle with any others but purely from pairs of particles~\cite{Yeo_2006}. For example, there is maximum entanglement between the following pairs: (a,b) and (c,d), and (a,c) and (b,d), and some non-maximal entanglement between (a,d) and (b,c). As with the product of EPR states, these states can be used to teleport an arbitrary two-qubit state from one place to another distant place and can also be used as resource of quantum dense coding~\cite{Yeo_2006, Lee_2002}. Moreover, as the sixteen $\chi$-type states form an orthonormal basis for the four-qubit space, they provide a new type of representation for four-qubit systems~\cite{Liu_2009}. 

The novel properties of the $\chi$ states led, naturally, to interest in how they might be prepared efficiently.  In 2009, Liu and Kuang proposed that these entangled states may be generated in four atomic qubits by employing the interaction between light and four atoms placed in four separate optical cavities~\cite{Liu_2009}.  In the same year, Wang and Zhang published a scheme with which to produce these states with a simple experimental setup  employing maximally and non-maximally polarization-entangled photons to encode the state of the output photons~\cite{Wang_2009}.The aim of our work is to propose an alternative scheme to generate the states $\ket{\chi^{ij}}$ by using two hyper-entangled photons, which are maximally entangled in both their polarization and orbital angular momentum.

The structure of this paper is as follows. \Sref{sec_2} introduces a required transformation that results in an entangled photon pair in the state $\ket{\chi^{00}}$, along with its corresponding quantum circuit.  In \sref{sec_3} we explain how each optical element affects the composite state, polarization and orbital angular momentum, of a light beam.  Finally, \sref{sec_4} presents our proposed optical system for implementing the quantum circuit and further transformations of the $\ket{\chi^{00}}$ to prepare any one of the states $\ket{\chi^{ij}}$.

\section{Transformation of two entangled photons}\label{sec_2}
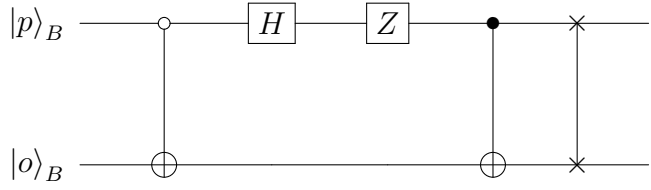
\begin{figure}
  \centerline{
	\Qcircuit @C=2.3em @R=3.5em {
	\lstick{\ket{p}_{B}}&\ctrlo{1}&\gate{H} &\gate{Z} &\ctrl{1}&\qswap&\qw\\
	\lstick{\ket{o}_{B}}&\targ&\qw&\qw&\targ&\qswap\qwx&\qw
	}
	}
 \caption{The quantum circuit corresponding to the transformation of the state of photon B given in~\eref{eq_7}, where $\ket{p}_{B}$ and $\ket{o}_{B}$ are the polarization and OAM parts of the composite state of the photon.\label{fig_2}}
\end{figure}

In~\cite{Barreiro_2005}, photons that are hyper-entangled in both their polarization and orbital angular momentum (OAM) are obtained by employing a 351 nm Argon ion laser with 120 mW power pumping two connected $\beta$-barium borate (BBO) crystals whose optical axes perpendicularly aligned. Photon pairs generated from the first crystal are horizontally polarized, while the second crystal produces the vertical polarization. As the non-linear crystals are in close proximity, the spatial modes of the output photons originating in each of the crystals are identical. Thus, the (unnormalized) state of the emitted photon pairs is 
\begin{equation}\label{eq_4}
(\ket{HH}+\ket{VV})^{p}_{AB}\otimes(\ket{RL}+\alpha\ket{GG}+\ket{LR})^{o}_{AB}
\end{equation}
where $H$ and $V$ denote horizontal and vertical polarizations, respectively, while $R$, $L$ and $G$ represent the modes with OAM $+\hbar$, $-\hbar$ and $0$ respectively for each photon\footnote{The precise form of this state contains also states with higher orbital angular momentum, but the relative sizes of these contributions can be controlled\cite{Alison,Alison2}.}. The superscripts $p$ and $o$ indicate the polarization and orbital angular momentum states, respectively, while the subscripts $A$ and $B$ indicate that it is the state of photon $A$ or $B$ respectively. The scalar quantity $\alpha$ is determined by mode-matching conditions. The photon pairs are also entangled in their emission times and frequencies, but in this work this type of entanglement
is important only in that it allows the use of arrival time at the detectors to select uniquely photon pairs that are entangled. 
\begin{figure}
\centering
\includegraphics[width=0.5\linewidth]{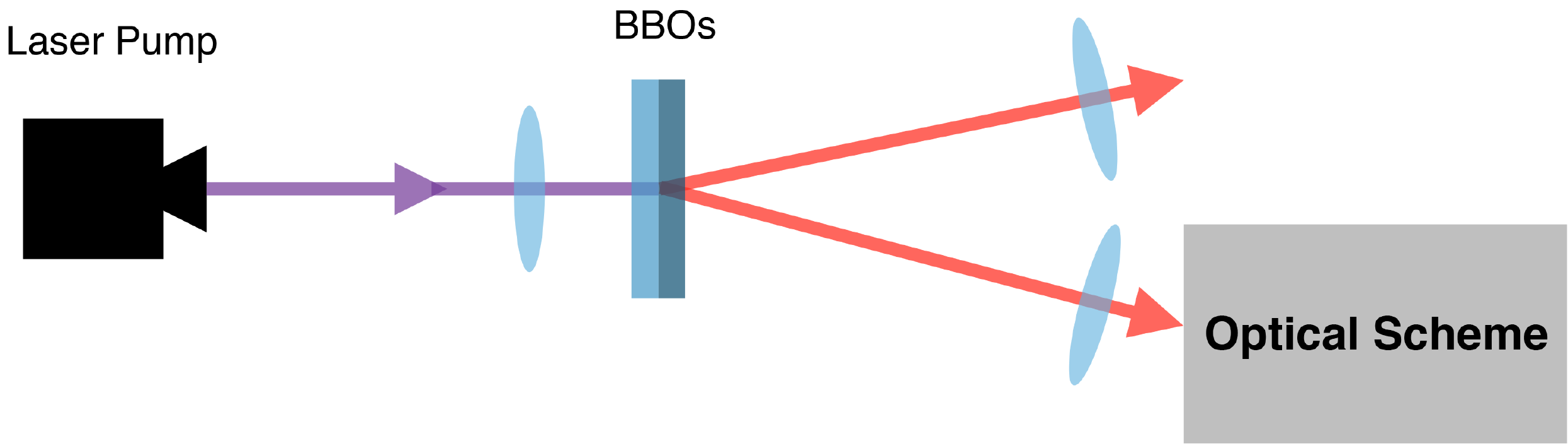}
\caption{The optical alignment to create hyper-entangled photon pairs by coherent sequential spontaneous parametric down-conversions (SPDC).\label{fig_1}}
\end{figure}
It is evident that we can obtain the maximally hyper-entangled state if the $\ket{GG}$ component is omitted.  This may be
achieved by spatial filtering to remove the beam centre or, more rigorously, by employing a mode-sorter 
\cite{Leach_2002} to select only odd-valued OAM states.  The selected state of the emitted photon pairs then becomes
\begin{eqnarray}\label{eq_5}
\ket{\Phi^{+}}^{p}_{AB} \otimes \ket{\Psi^{+}}^{o}_{AB}&=\frac{1}{2}(\ket{HH}+\ket{VV})^{p}_{AB} \otimes (\ket{RL}+\ket{LR})^{o}_{AB} \nonumber \\
&=\frac{1}{2}(\ket{00}+\ket{11})^{p}_{AB}\otimes(\ket{01}+\ket{10})^{o}_{AB}\label{eq_5b}
\end{eqnarray}
To obtain \eref{eq_5b} we encode $\ket{H}^p$ ($\ket{V}^p$) to be $\ket{0}^p$ ($\ket{1}^p$) and $\ket{R}^{o}$ ($\ket{L}^{o}$) to be $\ket{0}^{o}$ ($\ket{1}^{o}$). We can finally rewrite the state in terms of the superposition of the product states of the photons $A$ and $B$ as
\begin{equation}\label{eq_6}
\ket{X}_{AB}=\frac{1}{2}(\ket{00}_{A}\ket{01}_{B}+\ket{01}_{A}\ket{00}_{B}+\ket{10}_{A}\ket{11}_{B}+\ket{11}_{A}\ket{10}_{B}).
\end{equation}
The first qubits of photons $A$ and $B$, in this equation, now represent polarization states of these photons while the second represent their OAM states, so that, for example, $\ket{01}_{A}$ represents the composite state $\ket{H}^{p}\ket{L}^{o}$ of the photon $A$. We note the symmetric property of the photon pair: swapping the states of photons $A$ and $B$ leaves the state unchanged. To obtain $\ket{\chi^{00}}$, the state of one of the photons should be transformed as 
\begin{equation}\label{eq_7}
\eqalign{
\ket{00}_B\rightarrow\frac{1}{\sqrt{2}}(\ket{10}-\ket{01})_B,\cr
\ket{01}_B\rightarrow\frac{1}{\sqrt{2}}(\ket{00}-\ket{11})_B,\cr
\ket{10}_B\rightarrow\frac{1}{\sqrt{2}}(\ket{00}+\ket{11})_B,\cr
\ket{11}_B\rightarrow\frac{1}{\sqrt{2}}(\ket{10}+\ket{01})_B.}
\end{equation}
The above transformation is described by the quantum circuit shown in~\fref{fig_2}.
The first quantum gate in the circuit is the CNOT gate such that the target qubit will be flipped if the associated control qubit is $\ket{0}$.
If the swap gate is not included, we obtain
\numparts
\begin{eqnarray}\label{eq_8}
\ket{\chi^{00}_{\rm Lee}}=\frac{1}{\sqrt{2}}(\ket{\lambda^{0}}+\ket{\lambda^{1}})_{AB},\label{eq_8a}
\end{eqnarray}
with
\begin{eqnarray}
\ket{\lambda^{0}}\equiv\frac{1}{2}(\ket{0000}-\ket{0011}-\ket{0110}+\ket{0101})\label{eq_8b},\\
\ket{\lambda^{1}}\equiv\frac{1}{2}(\ket{1001}+\ket{1010}+\ket{1100}+\ket{1111})=\ket{\zeta^{1}}\label{eq_8c},
\end{eqnarray}
\endnumparts
which is the state $\ket{\chi^{00}}$ proposed by Lee {\it et al} in 2002~\cite{Lee_2002}. Including the swap gate at the end serves only to ensure that the OAM state of the photon B is realized to be the third qubit while its polarization is the last. After the given transformation, the state of the photon pairs turns out to be
\begin{equation}\label{eq_9}
\ket{\chi^{00}}=\frac{1}{\sqrt{2}}(\ket{\zeta^{0}}+\ket{\zeta^{1}})_{AB} ,
\end{equation}
as desired.

\section{Optical realization}\label{sec_3}
Before discussing the proposed optical system realizing the given transformation, we first explain how each of the optical elements we will employ act on the composite state of the photons. Let us start with birefringent wave plates, the effect of which on polarization of light is well known. Both half- and quarter- wave plates are wave retarders; their function is  to delay the phase of the polarization component lying in the direction of their slow axes by the phase $\pi$ and $\pi/2$ relative to the phase of the perpendicular component. The Jones matrices, which are used to describe effects of  optical elements on optical polarization, are for quarter and half wave plates:
\numparts
\begin{eqnarray}\label{eq_10}
J_{\rm q}(\theta)&=
\left(\begin{array}{cc}
   \cos \theta & -\sin \theta  \\
    \sin \theta  & \cos \theta
    \end{array}\right)
\left(\begin{array}{cc}
   1 & 0  \\
    0  & i
    \end{array}\right)
    \left(\begin{array}{cc}
   \cos \theta & \sin \theta  \\
    -\sin \theta  & \cos \theta
    \end{array}\right) \nonumber \\
&=\left(\begin{array}{cc}
    \cos^2\theta+i\sin^2\theta & (1-i)\sin\theta\cos\theta  \\
    (1-i)\sin\theta\cos\theta  & \sin^2\theta+i\cos^2\theta
    \end{array}\right),\label{eq_10b}\\
J_{\rm h}(\theta)&=
\left(\begin{array}{cc}
   \cos \theta & -\sin \theta  \\
    \sin \theta  & \cos \theta
    \end{array}\right)
    \left(\begin{array}{cc}
   1 & 0  \\
    0  & -1
    \end{array}\right)
    \left(\begin{array}{cc}
   \cos \theta & \sin \theta  \\
    -\sin \theta  & \cos \theta
    \end{array}\right) \nonumber\\
&=\left(\begin{array}{cc}
    \cos 2\theta & \sin 2\theta  \\
    \sin 2\theta & -\cos 2\theta
    \end{array}\right). \label{eq_10d}
\end{eqnarray}
\endnumparts
The angle $\theta$ gives the orientation, relative to the horizontal, of the fast axis~\cite{Fowles_1989, Hecht_1974}.
As these birefringent wave plates affect only the polarization of photons while leaving the OAM modes unaffected, the total effects of these wave plates on a composite state correspond to the transformation
\begin{equation}\label{eq_11}
\ket{p^\prime,o^\prime}=J_{i}\otimes I\ket{p,o} \quad \textrm{with} \quad i={\rm q},{\rm h},
\end{equation}
where $\ket{p,o}$ and $\ket{p^\prime,o^\prime}$ are the composite states of a photon before and after passing through the optical elements.

\begin{figure}
\centering
\includegraphics[width=0.5\linewidth]{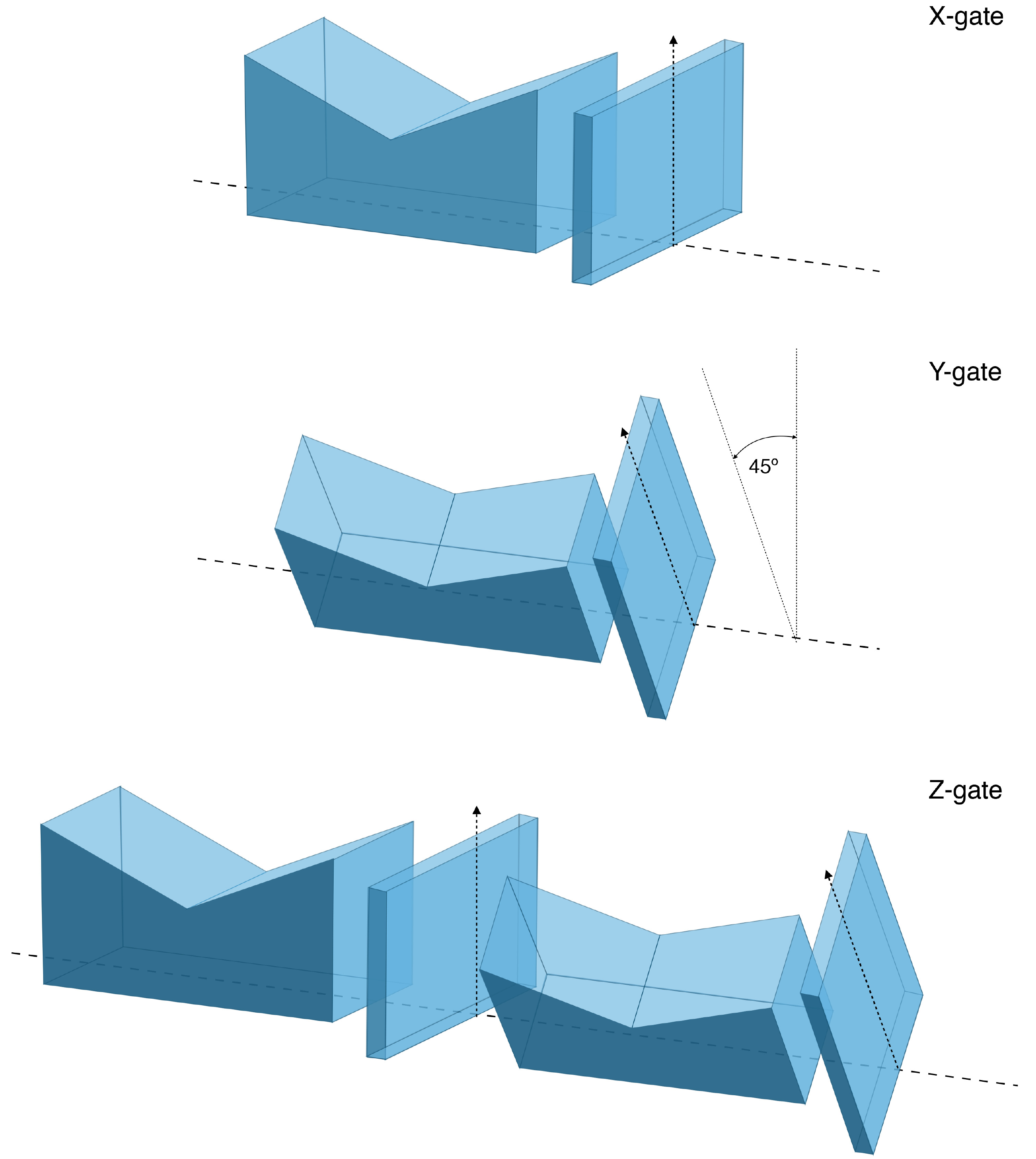}
\caption{The particular alignments of M-shaped Dove prisms together with quarter-wave plates, to compensate their polarization effect, which provide physical realizations of the Pauli gates for OAM qubits when $l=\pm1$. The dashed arrows represent the fast axes of the quarter-wave plates.\label{fig_4}}
\end{figure}

A Dove prism is an optical element frequently employed in optical orbital angular momentum experiments, which acts to flip the sign of the orbital angular momentum of light. For example, in this work  it converts the OAM mode from $l=1$ to $-1$ or vice versa. A Dove prism, whatever shape it is, was originally invented to invert an image, which means if it is included into the path of a light beam, it will generally give some reflection and refraction to the beam. As, according to the Fresnel equations, both reflection and refraction introduce change in beam polarization, a Dove prism has a particular effect on both polarization and OAM modes of an incident beam depending on its individual shape.  In this work we consider M-shaped Dove prisms, which are shown in \fref{fig_4}, as their Jones matrix is rather simple. Indeed this type of Dove prism was originally invented to be a quarter wave retarder~\cite{Bennett_1970}; thus its effect on polarization is simply the same as a quarter-wave plate, with Jones matrix:
\begin{equation}\label{eq_12}
J_{\rm D}=\left(\begin{array}{cc}
    1 & 0  \\
    0 & i
    \end{array}\right).
\end{equation}
Therefore, we can describe its effect on a composite state by
\begin{equation}\label{eq_13}
\ket{p^\prime,o^\prime}=J_{\rm D}\otimes\sigma^{x}\ket{p,o}.
\end{equation}
As a result, if one wishes only to invert the OAM state of an incident beam while leaving its polarization unaffected, one
needs to add a quarter-wave plate before or after the Dove prism to compensate its phase retardation effect.

Another optical element that will be discussed is a polarizing beam splitter (PBS). Its function is rather obvious in its name as it splits an optical beam into two different distinct beams whose travelling paths depend on the optical polarization of the beam. In this work all PBSs are considered to transmit horizontal and reflect vertical polarized beams. Therefore, we can treat each type of optical beams separately, and it allows us to realize control gates for composite states such that the polarization and OAM states are the control and the target qubits respectively. 

\begin{figure}
\centering
\includegraphics[width=0.5\linewidth]{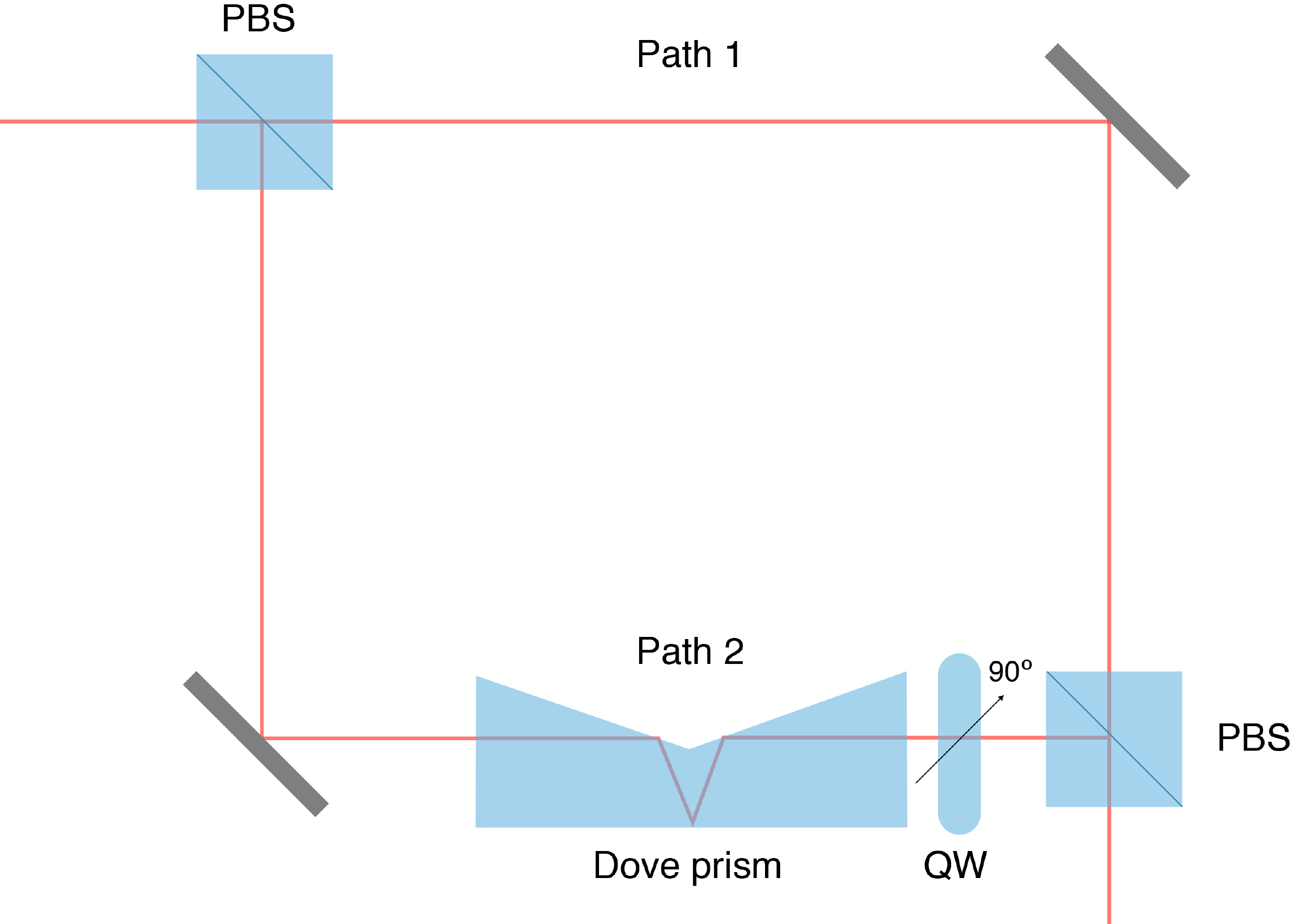}
\caption{This figure shows the interferometer that realizes the CNOT gate for the composite state by adding an M-shaped Dove prism and a quarter-wave plate with its fast axis at angle $\pi/2$ with respect to the horizontal plane into path 2. Alternatively, the same result can be given without the quarter-wave plate if both internal paths are adjusted such that the relative phase shift between different polarizations becomes a global phase shift of photons. \label{fig_3}}
\end{figure}

As an illustration of the operation of these devices, let us consider the interferometer given in~\fref{fig_3}.  The first PBS separates the incident beam into two different paths. The horizontally and vertically polarized beams travel along the internal paths 1 and 2 of the interferometer respectively, and  they are combined again at the second PBS. Therefore, if we introduce a Dove prism together with a quarter-wave plate into one of these paths, the overall effect will be to flip the OAM state of the composite system for just one component of polarisation - that corresponding to the path in which these optical elements are put. For example, in~\fref{fig_3}, we put a Dove prism and a quarter-wave plate into the path 2, so that if the incident beam is horizontally polarized, it will be forced to travel along the path 1 and not encounter any elements affecting its composite state, and its composite state is untouched. On the other hand, if the beam is vertically polarized, it will be only permitted to go along the internal path 2 and encounter both the M-shaped Dove prism and the quarter-wave plate. The OAM part of the composite state is flipped in this case. This means that for a composite state of polarization and OAM the interferometer given in this figure acts as a CNOT gate such that if the polarization qubit, the control qubit in this case, is $\ket{1}^{p}$  the OAM qubit will be inverted, while it is left unchanged if the control qubit is $\ket{0}^{p}$.  Devices built on this principle 
have been used, successfully, to measure both the OAM and the spin for light at the single photon level
\cite{Leach_2002,Wei_2003,Leach_2004}.

At this point, one can notice that, according to \eref{eq_10d}, single-qubit gates, such as the Pauli gates and Hadamard gates, for polarization can be realized by appropriately orientated half- wave plates. For example, the realizations of the Z- and X-gates are just  half-wave plates having fast axes parallel to and oriented at an angle $\pi/4$ with respect to the horizontal plane. The Y-gate of polarization qubits can be obtained by using the fact that $\sigma^{z}\sigma^{x}=\rmi \sigma^{y}$ which means we have to use two half-wave plates with different orientation to realize the Y-gate.

The complex amplitude of a Laguerre-Gaussian (LG) beam has the phase dependent term, $\exp({\rm i} l\phi)$ where $l$ is the orbital angular momentum quantum number of the light beam~\cite{Allen_1992}. Thus when one rotates the beam by an angle $\alpha$ this term will become $\exp({\rm i} l(\phi+\alpha))$, or in other words rotation of an LG beam contributes a phase shift  of $\Delta \psi=l\alpha$~\cite{Courtial_1998, Leach_2002}. An optical beam can be rotated by suitably oriented Dove prisms. A non-rotated Dove prism gives a non-rotated, reflected image, and when the Dove prism is rotated by an angle $\beta$, the reflected image is rotated through $2\beta$. Non-rotated Dove prisms act as an X-gate for OAM qubits as they change the OAM state of an optical beam to be the opposite state, from $\ket{l}$ to $\ket{-l}$. The single-qubit Y- and Z-gates for OAM can be implemented as follows: a Dove prism rotated by $\pi/4$ with respect to the vertical plane transforms an OAM qubit as
\begin{equation}\label{eq_14}
\eqalign{
\ket{0}^{o}&\rightarrow \rme^{-\rmi\pi/2}\ket{1}^{o},\\
\ket{1}^{o}&\rightarrow \rme^{\rmi\pi/2}\ket{0}^{o},}
\end{equation}
when $l=\pm1$. This is exactly the transformation corresponding to acting with a Y-gate on a qubit. As we know that $\sigma^{x}\sigma^{y}={\rm i}\sigma^{z}$, one can implement the Z-gate of OAM qubits by using two Dove prisms: non-rotated and rotated by $\pi/4$ Dove prisms, respectively. We must also remember to take account of the fact that
each Dove prism has an effect on the polarization, as discussed above. The total effect of the rotated Dove prism at angle $\theta$ on a composite qubit can thus be written as
\begin{equation}\label{eq_15}
\ket{p^\prime,o^\prime}=J_{\rm q}(\theta)\otimes \left(\begin{array}{cc}
    0 & \rme^{2\rmi\theta}  \\
     \rme^{-2\rmi\theta} & 0
    \end{array}\right)\ket{p,o}.
\end{equation}
 To compensate for this a quarter-wave plate should be added before or after each Dove prism. The physical realizations of the Pauli gates are depicted in~\fref{fig_4}.

All effects of optical elements we discussed above on composite qubits are summarized in \tref{tab_1}.
\begin{table}[hb]
\caption{\label{tab_1}Summary of the effects of optical elements on composite qubits}
\begin{tabular}{@{}*{7}{l}}
\br
Optical Elements&Effects on composite qubits\\
\mr
Quarter-wave plate fast axis at angle $\theta$& $U_{\rm q}(\theta)=J_{\rm q}(\theta)\otimes I$\\
Half-wave plate with fast axis at angle $\theta$& $U_{\rm h}(\theta)=J_{\rm h}(\theta)\otimes I$\\
Rotated M-shaped Dove prism at angle $\theta$& $U_{\rm D}(\theta)=J_{\rm q}(\theta)\otimes \left(\begin{array}{cc}
    0 & \rme^{2\rmi\theta}  \\
     \rme^{-2\rmi\theta} & 0
    \end{array}\right)$\\
\br
\end{tabular}
\end{table}

\section{Optical system}\label{sec_4}
In this section, we seek an optical system that transforms the composite state of the photon B so as to prepare photons A and B in any one of the entangled states $\ket{\chi^{ij}}$.  We start with the optical transformation corresponding
to the quantum circuit in~\fref{fig_2}.  This quantum circuit includes two different CNOT gates, two single-qubit gates, Hadamard and Y-gates, and the swap gate at the end.  We use optical elements corresponding to each of these.

The principle of the first CNOT gate in the circuit is that if the control qubit, the polarization qubit, is in the state 
$\ket{0}^{p}$, then the target qubit, the OAM qubit, will be flipped, while it is left unchanged if the control qubit is $\ket{1}^{p}$. As mentioned in the previous section, the CNOT gates in the quantum circuit can be realized by interferometers with a M-shaped Dove prism in one of their internal paths. Recall that the logical qubits, $\ket{0}^{p}$ and $\ket{1}^{p}$, are encoded as horizontal and vertical polarization states, $\ket{H}$ and $\ket{V}$, respectively. This means we can realize the first CNOT gate by the same interferometer as given in~\fref{fig_3}, but the Dove prism must be in the internal path 1, the path associated with the horizontally polarized component of the input beam, rather than path 2.
As the Dove prism leaves horizontal polarization unaffected, we do not need to add a quarter-wave plate in this case. The second CNOT gate is slightly different from the first one as it will flip the target qubit if the control qubit is in the state $\ket{1}^{p}$, and do nothing if the control qubit is $\ket{0}^{p}$. This means the second CNOT gate can be exactly realized as the interferometer in~\fref{fig_3}. 

According to~\eref{eq_10d}, we can explicitly see that the Jones matrices of half- wave plates with optical axes in the horizontal plane and at the angle $\pi/8$ are exactly the matrix representations of the Pauli Z and Hadamard gates respectively. This implies the Hadamard and Z gates in the circuit can then be realized as appropriately oriented half-wave plates.

\begin{figure}
\centering
\includegraphics[width=0.7\linewidth]{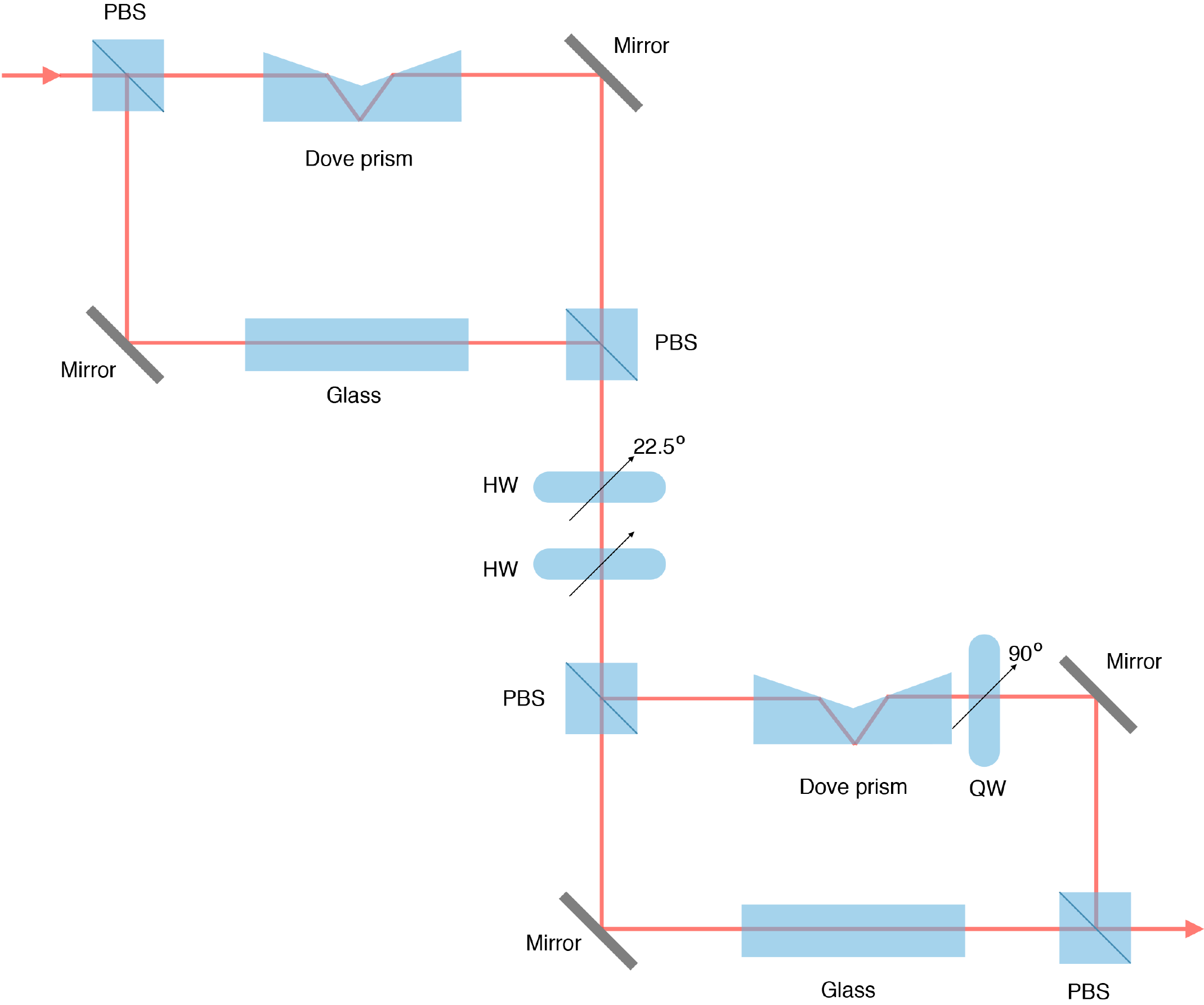}
\caption{This figure shows our proposed optical system corresponding to the quantum circuit given in~\fref{fig_2}. The lengths of the pieces of glass in the two interferometers are selected so as to compensate for the delay associated with propagation through the Dove prisms. The quarter wave plate in the second interferometer can be removed if the interferometer path length is suitably adjusted.  \label{fig_5}}
\end{figure}

Recall that even without the swap gate at the end we still arrive at a realization of the state $\ket{\chi^{00}}$. 
The swap gate of the quantum circuit can be implemented easily by relabelling the composite state of photon B as mentioned earlier. With this swap gate, Yeo's version of the $\ket{\chi^{00}}$ state is finally obtained. The optical system corresponding to the quantum circuit given in~\fref{fig_2}  is illustrated in~\fref{fig_5}.

With \eref{eq_3b}, once we obtain $\ket{\chi^{00}}$, any of the 15 other $\chi$-type four-party entangled states can be generated using only local Pauli operations:
\begin{equation*}
\ket{\chi^{ij}}=\sigma^{i}\otimes\sigma^{j}\otimes I\otimes I\ket{\chi^{00}}_{AB}.
\end{equation*}
At this point our task is to implement Pauli gates for both polarization and OAM states. As mentioned in~\sref{sec_3}, the Pauli gates of polarization and OAM qubits can be implemented by half-wave plates and Dove prisms (together with quarter-wave plates to compensate the polarization effect of these Dove prisms) respectively. For example, the tensor product of the Pauli X and Y operators for polarization and OAM qubits, respectively, $\sigma^{x}_{p}\otimes\sigma^{y}_o$, can be realized by a half-wave plate with fast axis at angle $\pi/4$ and a $\pi/4$-rotated M-shaped Dove prism together with a quarter-wave plate with its fast axis at $\pi/4$ with respect to the horizontal plane. It follows that 
an arbitrary state $\ket{\chi^{ij}}$ in the set of $\chi$-type states can be realized by applying birefringent wave plates and M-shaped Dove prisms with specified orientations after the optical system shown in~\fref{fig_5}.

\section{Conclusion}
We have presented the operation required to transform the maximally hyper-entangled state of a photon pair, obtained from SPDC process, into $\ket{\chi^{00}}$, a state with genuine four-party entanglement. We have shown the effect of each optical element we use on the composite state, and an optical system suitable for preparing $\ket{\chi^{00}}$ has been proposed. To obtain any other of the $\chi$-type states $\ket{\chi^{ij}}$, further simple transformations are required and these may be realized using birefringent wave plates and M-shaped Dove prisms.

As the proposed optical system requires only readily available linear optical components, preparation of the desired
states should be possible using current technology.  Unlike previous work the proposed scheme does not require any post selection, so the efficiency of successful transformation does not depend, intrinsically, on the efficiency of photon detectors. We hope that our scheme for the production of elements of this class of multipartite entangled states may be realized experimentally and that doing so will give us better insight of multipartite entanglement, and enable the demonstration of novel quantum
information protocols.

\ack
A Ritboon and S M Barnett acknowledge support from the Development and Promotion of Science and Technology Talents Project (DPST), Thailand, and the Royal Society (RP150122) respectively.

\end{document}